\documentclass[12pt,a4paper]{article}

  \usepackage{a4wide}
  \usepackage{latexsym}
  \usepackage{epsf}
  \usepackage{amssymb}
  \usepackage{graphicx}
  \usepackage{amsmath, cite}
  \usepackage{amsmath,amssymb,amsthm}
  \usepackage{verbatim}
  \usepackage{hyperref}

\renewcommand{\d}{\textrm{d}}
\newcommand{\e}{\textrm{e}}

\newcommand{\be}{\begin{equation}}
\newcommand{\ee}{\end{equation}}
\newcommand{\ba}{\begin{eqnarray}}
\newcommand{\ea}{\end{eqnarray}}

\renewcommand{\d}{\textrm{d}}

\begin{document}
\numberwithin{equation}{section}

\begin{center}

\begin{flushright}
{\small UUITP-02/13
 \\ IPhT-T13/009
} \normalsize
\end{flushright}
\vspace{0.3 cm}

{\LARGE \bf{Antibranes don't go black\\\vspace{0.2cm}}}

\vspace{1.1 cm} {\large I. Bena$^a$, J. Bl{\aa }b\"ack$^b$,  U.H. Danielsson$^b$, T. Van
Riet$^c$}\\

\vspace{0.8 cm}{$^{a }$Institut de Physique Th\'eorique, CEA
Saclay, \\CNRS URA 2306 ,  F-91191 Gif-sur-Yvette, France}\\
\vspace{0.1 cm}{$^{b}$ Institutionen f{\"o}r fysik och astronomi,\\ Uppsala Universitet, Uppsala, Sweden}\\
\vspace{.1 cm} {$^c$ Instituut voor Theoretische Fysica, K.U. Leuven,\\
Celestijnenlaan 200D B-3001 Leuven, Belgium} \footnote{{\ttfamily {iosif.bena @ cea.fr,   johan.blaback,  ulf.danielsson @ physics.uu.se, thomasvr @ itf.fys.kuleuven.be}}}\\

\vspace{0.8cm}

{\bf Abstract}
\end{center}

\begin{quotation}
When D-branes are inserted in flux backgrounds of opposite charge, the resulting solution has a certain singularity in the fluxes. Recently it has been argued, using numerical solutions, that for anti-D3 branes in the Klebanov-Strassler background these singularities cannot be cloaked by a horizon, which strongly suggests they are not physical. In this note we provide an analytic proof that the singularity of all codimension-three antibrane solutions (such as anti-D6 branes in massive type IIA supergravity or anti-D3 branes smeared on the $\mathbb{T}^3$ of  $\mathbb{R}^3\times \mathbb{T}^3$ with fluxes) cannot be hidden behind a horizon, and that the charge of black branes with smooth event horizons must have the same sign as the charge of the flux background. Our result indicates that infinitesimally blackening the antibranes immediately triggers brane-flux annihilation, and strengthens the intuition that antibranes placed in flux with positive charge immediately annihilate against it.

\end{quotation}
\newpage

\section{Introduction}

Finding controllable and explicit ways for supersymmetry breaking in string theory compactifications remains an important challenge. One possibility is offered by inserting brane sources with opposite orientation to some background flux solution; if the flux background is mutually-BPS with certain D-branes then adding D-branes with opposite charge breaks supersymmetry. If one furthermore inserts the anti-branes  in regions of very high space-time warping, one can tune down the supersymmetry-breaking energy way below the string scale, and use this technique to obtain de Sitter vacua and inflation in string theory \cite{Kachru:2003aw, Kachru:2003sx}, to engineer holographically supersymmetry breaking in gauge theories \cite{Kachru:2002gs} or to construct very large families of microstate geometries (or fuzzballs) of non-extremal black holes \cite{Bena:2011fc, Bena:2012zi}. 

In recent years this picture has been questioned because the supergravity solutions that should describe these configurations have certain 
singularities in the fluxes, that appear to be unphysical. The typical example, corresponding to the insertion of anti-D3 branes in the Klebanov-Strassler throat solution \cite{Klebanov:2000hb}, gives supergravity solutions \cite{Bena:2009xk, Bena:2011wh , Bena:2011hz, Massai:2012jn, Bena:2012bk, Gautason:2013zw} (see \cite{DeWolfe:2008zy, McGuirk:2009xx } for earlier work) with singular three-form fluxes, that do not have the right orientation to correspond to any type of brane. One might have thought that this flux singularity signifies the intention of the anti-D3 branes to polarize into meta-stable D5 branes, as it happens in the probe approximation \cite{Kachru:2002gs}, but, as it has been argued in \cite{Bena:2012vz}, this does not appear to happen: one can calculate the polarization potential {\`a la Polchinski-Strassler}, and show that even if this potential has all the terms one expects from  \cite{Polchinski:2000uf}, the exact coefficients of these terms are such that the polarization is impossible. 

 There is a simple physical picture that provides a possible explanation of the three-form singularity and the breakdown of the probe approximation \cite{Blaback:2010sj,Blaback:2011nz,  Blaback:2012nf}. In simple words this picture regards the background fluxes as a delocalized fluid of D3 branes (see also \cite{DeWolfe:2004qx}). The three-form singularity is then nothing else but the unphysical piling up of D3 branes (dissolved in fluxes) near the anti-D3 brane, caused by the combined gravitational and electromagnetic attraction. If this attraction could be balanced by the internal pressure of the fluid of D3 branes, the supergravity solution would describe a finite flux clumping. However, one finds from the exact solution that this clumping is infinite, which suggests that the singularity signals that the true solution wants to be time-dependent. Further evidence for this interpretation was presented in \cite{Blaback:2012nf}, where the effect of the flux piling was qualitatively evaluated in the probe action. The outcome is that flux piling facilitates the annihilation of the anti-D3 branes with the surrounding flux, which could have been intuitively expected from the fact that branes and antibranes annihilate easier when they are closer. 

Besides checking whether the obvious resolution mechanisms save this singularity, one can take an alternative route and be agnostic about the mechanism, but argue that string theory will somehow resolve singularities that satisfy certain criteria, and hence will also resolve this one. The most popular singularity-resolution criterion was formulated in \cite{Gubser:2000nd}, where it was argued that singularities that can be cloaked by a regular event horizon are resolvable in string theory. This criterion has never failed until now\footnote{In contrast, the criterion formulated in \cite{Dymarsky:2011pm}, that singularities with a finite action should be acceptable in string theory fails dramatically when applied to the negative-mass Schwarzschild black hole singularity, which must clearly be unphysical and not resolvable in any quantum theory of gravity \cite{Horowitz:1995ta}.}, and correctly post-dicts the resolution \cite{Polchinski:2000uf} of the GPPZ singularity \cite{Girardello:1999bd}  from the existence of the Freedman-Minahan black hole \cite{Freedman:2000xb}, or the resolution \cite{Klebanov:2000hb} of the Klebanov-Tseytlin (KT) singularity  \cite{Klebanov:2000nc} from the existence of the KT black hole   \cite{Buchel:2000ch, Buchel:2001gw, Gubser:2001ri, Aharony:2007vg, Buchel:2010wp}.

If one tries to apply this resolution criterion to the singularity of antibranes in the Klebanov-Strassler (KS) solution one would expect that this singularity is physical if these antibranes can be cloaked by a horizon, or alternatively if there exists a KS or a KT black hole with a charge opposite to that of the background. However, this turns out not to happen: the black holes in KT and KS, both regular and mass-deformed, have been constructed numerically by \cite{Aharony:2007vg} and \cite{Buchel:2010wp}, and, throughout the parameter space explored in the numerical construction, the charge at the horizon of these black holes has always the same sign as the asymptotic charge of the solution \cite{Bena:2012ek}. Hence, it is very unlikely that such black holes exist, which indicates that the singularity of antibranes is unphysical. 

The absence of KS and KT black holes with negative charges, which was  discovered numerically in \cite{Aharony:2007vg,Buchel:2010wp,Bena:2012ek} cries out for an analytic understanding, and it is our purpose in this paper to provide one in a simplified setup. We consider anti-D6 branes in a massive type IIA supergravity background with D6 brane charge dissolved in the fluxes, as well as its T-duals, that include for example anti-D3 branes smeared on the $\mathbb{T}^3$ of  a $\mathbb{R}^3\times \mathbb{T}^3$ with positive D3 charge dissolved in the fluxes.  The solution describing these  anti-D3 branes \cite{Blaback:2011nz, Blaback:2011pn} has been shown to capture the essential features of the anti-D3 brane KS solution: First, the singularity of the KS perturbative anti-D3 solution  \cite{Bena:2009xk} has multiple singular three-form components \cite{Massai:2012jn}, but the fully backreacted solution  \cite{Bena:2012bk} only has one singular component\footnote{The fact that the fully backreacted KS antibrane solution will have less singular fluxes than the perturbative one
was first argued in \cite{Dymarsky:2011pm}; however, as intuited in \cite{Massai:2012jn}, the mechanism discussed there only removes two of of the three singular flux components.}, which is identical to that of the anti-D3 branes on $\mathbb{R}^3\times \mathbb{T}^3$. Second, the Polchinski-Strassler  potential for the anti-D3 branes to polarize into D5 branes computed in the $\mathbb{R}^3\times \mathbb{T}^3$ model  \cite{Bena:2012tx} captures three of the four terms of the Polchinski-Strassler  potential for the anti-D3 branes in KS  \cite{Bena:2012vz}. The extra term of the KS potential only makes polarization less likely, which indicates that the antibrane KS singularity is - if anything - worse than the antibrane singularity in $\mathbb{R}^3\times \mathbb{T}^3$.

In this paper we prove analytically that the anti-D6 singularity as well as its cousin anti-D3 brane singularity in $\mathbb{R}^3\times \mathbb{T}^3$ cannot be cloaked by a horizon by showing that the candidate black antibrane solutions must have singular three-form fluxes at the would-be horizon. The only possible black brane solutions must have the same sign of the charge at the horizon as the sign of the charge dissolved in the background fluxes. Our result confirms analytically the physics behind the numerical results of  \cite{Aharony:2007vg,Buchel:2010wp,Bena:2012ek}, and we hope will provide the starting point in a programme to prove analytically that no KS and KT black holes exist. 

In the next section we derive our result for the anti-D6 black brane solution using exactly the same technique as in \cite{Blaback:2011nz}. In section  \ref{T-dual}  we generalize this calculation to the smeared anti-D$p$ brane singularities with $p<6$ that come from the T-dualization of the anti-D6 singularity, and show that these singularities also cannot be cloaked by an event horizon. In section \ref{Spec} we end with some speculative interpretation of our results and point out an interesting similarity with black holes in asymptotically Lifshitz spacetimes \cite{Danielsson:2009gi}.

\section{The black (anti-)D6 brane}

In this Section we describe the ansatz for a black (anti-) D6 brane in a background with non-zero $H$ and $F_0$ fluxes and analyze the equations of motion. To describe a black 'anti-brane' we demand that the charge measured at the horizon and the asymptotic charge can have opposite signs. The only difference from the previous investigations of anti-D6 branes \cite{Blaback:2011nz, Blaback:2011pn} is the presence of a `blackening' factor in the ansatz, which in the absence of fluxes determines where the black hole horizon is. It turns out that the technical steps of the derivation of the singularity of anti-D6 branes  \cite{Blaback:2011nz} can be easily generalized to a proof that all putative black anti-D6 brane solutions must have a singular horizon. 


Upon  fixing the gauge for the $r$ redefinitions, the Einstein-frame ansatz is given by
\begin{eqnarray}
\label{TheBackground} 
\d s^2_{10} & = &
   \e^{2 A(r)}\Bigl(- \e^{2f(r)}\d t^2 + \d x_6^2 \Bigr) +\e^{2B(r)}\Big( \e^{-2f(r)}\d r^2 + r^2 (\d \psi^2 + \sin^2 \!\psi\, \d \phi^2 )\Big)\, , \\
H_3 & = & F_0 \lambda(r)  \e^{\tfrac{7}{4}\phi(r) + 3B(r)-f} r^2 \d r \wedge \omega_{S^2} \, , \qquad F_0  \neq  0, \ \\
F_2 & = & \e^{-\tfrac{3}{2}\phi(r) - 7 A(r) + B(r)} \alpha'(r) r^2 \,\omega_{S^2}   \, .
\end{eqnarray}
We find this gauge useful because it connects to the usual expressions for the branes in flat space (without extra fluxes). The extremal solution, with $f=0$, was found in \cite{Janssen:1999sa}, but we do not need its explicit form here.

The standard black D6 brane solution in flat space (without the $H$ and $F_0$ flux) has 
\be
e^{2f} = 1 -\frac{k}{r}\,.
\ee
with the horizon at $r=k>0$. In the presence of non-zero fluxes $H, F_0$ the solution is going to be more involved and we only demand that near the horizon the blackening factor approaches zero while the other functions remain finite, exactly as for the flat-space black hole.

\begin{figure}[h!]
\centering
\includegraphics[scale=0.5]{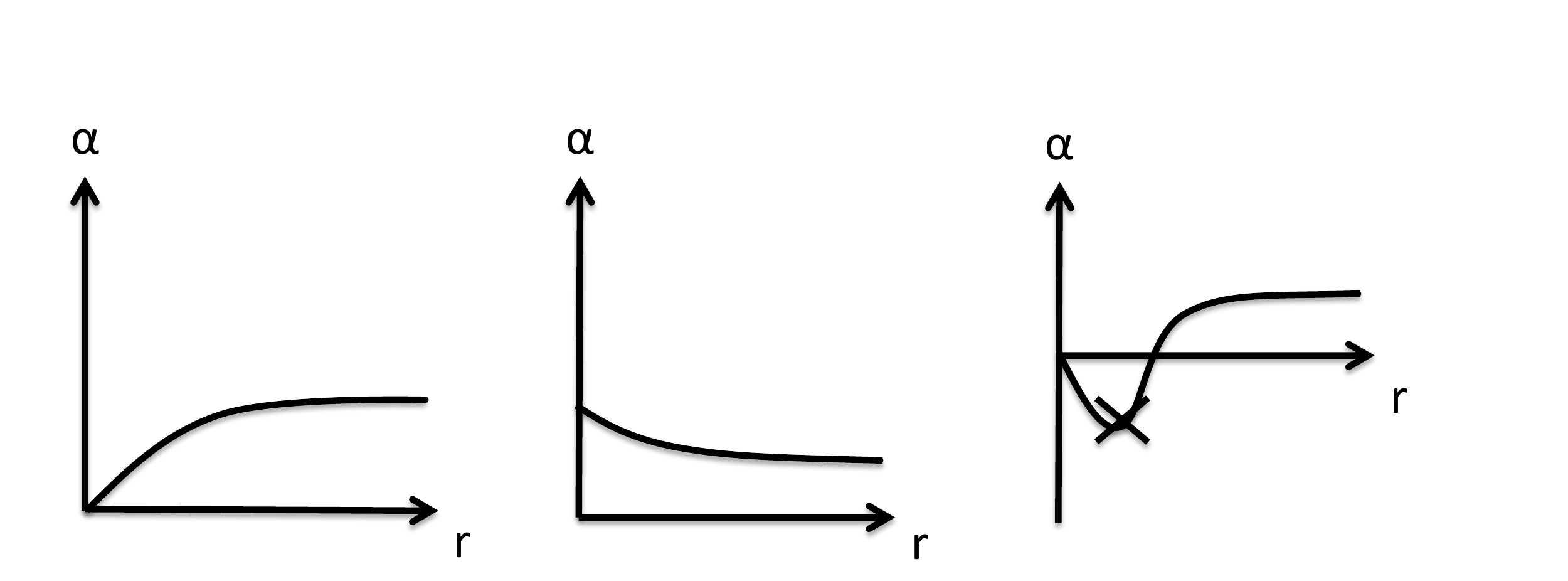}
\caption{\small{ \emph{ The possible $\alpha$ profiles for solutions that have positive charge dissolved in fluxes at large radius (positive $\alpha$). Only when the derivative of $\alpha$ is positive at the origin a curve exists that vanishes at the origin (left plot). Curves with negative derivative at the origin necessarily must have some non-zero value for $\alpha$ at the origin (middle plot). If one insists on $\alpha$ having a negative derivative at zero, then it must have a turnaround point at some radius where $\alpha$ is negative, which is forbidden by Eq. (\ref{topological}), as shown in the right plot.}}\label{boundary}}
\end{figure}

The essential information that leads to the no-go theorem for the black anti-brane can be derived solely from the $H$ equation of motion and the $F_2$ Bianchi identity
\begin{align}
& \d (\star_{10}\e^{-\phi} H) = -\e^{\tfrac{3}{2}\phi} F_0\star_{10}F_2\,,\\
& \d F_2 = F_0 H + Q\delta\,.
\end{align}
 The EOM for $H$ relates $\lambda$ and $\alpha$:
\be \label{algebraic}
\alpha = \lambda \e^{\tfrac{3}{4}\phi +7A +f}\,.
\ee
Together with the Bianchi identity we can derive the following second--order differential equation for $\alpha$, away from the delta function source:
\be \label{topological}
\boxed{
\e^{-\tfrac{3}{2}\phi -7A -B }\,\,\alpha'' + \frac{(\e^{-\tfrac{3}{2}\phi-7A +B}r^2)'}{\e^{3B}r^2}\,\,\alpha' = \e^{\phi-7A -2f}F_0^2 \,\,\alpha\,.
}
\ee
Apart from an extra factor of $\e^{-2f}$ on the left hand side, this equation is identical to the one derived in \cite{Blaback:2011nz}.
This equation fixes the rough behavior (`topological behavior') of the function $\alpha$ as follows: Near the origin (horizon) the sign of the charge has the same sign as the derivative of the `vector potential' $\alpha$ since, up to normalizations,
\be
Q = \alpha' \, r^2\,\e^{-\tfrac{3}{2}\phi -7A +B}|_{\text{horizon}}\,.
\ee
The sign of the charge dissolved in fluxes equals the sign of $\lambda$, which in turn equals the sign of $\alpha$ as a consequence of (\ref{algebraic}).  Equation (\ref{topological}) implies then that at a possible turning point of the $\alpha$ curve ($\alpha'=0$) the second derivative must have the same sign as the function itself. This implies that the function $\alpha$ takes a \emph{non-zero} and positive value at the black brane horizon $r=k>0$ if one insists that the black hole charge $Q$ and the asymptotic charge have opposite sign.  This is illustrated in figure \ref{boundary}. 
It is now straightforward to demonstrate that a non-zero value for $\alpha$ at the horizon implies a singular $H$ field strength at the horizon.  A simple computation shows that 
\be \label{H^2}
\e^{-\phi}H^2 = \alpha^2 \e^{\phi -14A -2f}F_0^2\,.
\ee
 
Since $e^{-2f}$ blows up at the horizon and since all the other fields have finite values, this indicates that we have a divergent energy density at the horizon and that this horizon is not smooth. 

\section{T-duality and simplified KS-like geometries}\label{T-dual}
We can T-dualize the anti-D6 Ansatz to obtain black $p$-brane Ansatze, for $p<6$  that describe branes smeared over the T-duality circles. This leads to Klebanov-Strassler-like warped geometries that have a base of the form:
\be
\mathbb{R}^3\times \mathbb{T}_{6-p}\,.
\ee
Since we have broken the D6 brane Lorentz isometry, we can consider a metric Ansatz that is a bit more general:
\be
\d s^2_{10} = e^{2 A(r)} ( - e^{2f(r)}\d t^2 + \d x^2_{p} ) + e^{2B(r)} (e^{-2f(r)}\d r^2 + r^2 \d \Omega_2^2 + \e^{2C(r)}\d \mathbb{T}_{6-p}^2)\,.
\ee
where $ \d \mathbb{T}_{6-p}^2$ is the metric on the $(6-p)$-dimensional torus built from the T-duality circles. The torus is filled with RR flux 
\be
F_{6-p}=m \,\epsilon_{6-p}\,,
\ee
where $m$ is a constant and $\epsilon_{6-p}$ the volume form on the torus (without the conformal factor $e^B$). The $H$-field strength fills the non-compact $\mathbb{R}^3$:
\be\label{hodge}
 \star_{9-p}H = \lambda(r) \e^{\frac{p+1}{4}\phi(r) } F_{6-p}\,.
\ee
Finally the $F_{8-p}$ Ansatz is  given in terms of a one-form $\d\alpha$, which points in the radial directions since $\alpha$ only depends on $r$:
\be
F_{8-p} = \e^{-(p+1)A-\frac{p-3}{2}\phi - f}\star_{9-p} \d \alpha\,.
\ee
In the above equations the functions $A, B, \lambda, f, \alpha$ depend on $r$ only  and the hodge stars have all metric functions included. One can readily check that for $p=6$ all the expressions in this section reduce to the expressions of the previous section. When $p=3$ the solution describes D3 branes with transverse space $\mathbb{R}^3\times \mathbb{T}_3$, and its near-D3 limit is similar to the infrared of a deformed conifold solution with smeared branes on the conifold $S^3$: 
as long as the curvature of the D3 branes dominates the curvature of the $S^3$, the $S^3$ and $S^2$ in $T^{1,1}$ can be approximated respectively by $\mathbb{T}_3$ and the $S^2$ in $\mathbb{R}^3$. The extremal solutions, are characterized by constant $\lambda$
\begin{equation}
\lambda =\pm 1\,.
\end{equation}
For $p=3$ this reproduces the well known (anti-) imaginary self duality condition on the three-form fluxes, through equation (\ref{hodge}). The extremal solutions are however not SUSY. For $p=3$ this can be traced back to the fact that the three-form fluxes have a non-zero $(0,3)$ component \cite{Grana:2000jj}. Supersymmetry can however be restored by adding worldvolume dependence to the warp factors \cite{Janssen:1999sa, Blaback:2012mu}.

The relevant equations for proving the no-go theorem for black anti-branes are again the $H$ equation of motion and the $F_{8-p}$ Bianchi identity
\begin{align}
& \d (\star_{10} \e^{-\phi} H) = -\e^{\tfrac{p-3}{2}\phi}\,\star_{10}F_{8-p}\wedge F_{6-p}\,,\\
& \d F_{8-p} = H \wedge F_{6-p}\,.
\end{align}
The $H$ equation of motion relates $\alpha$ and $\lambda$ as follows
\begin{equation}
\alpha=\lambda \e^{(p+1)A+\frac{p-3}{4}\phi +f} \,.
\end{equation}
Combining this with the  $F_{8-p}$ Bianchi identity, away from the source, we find
\begin{equation}
\boxed{
\e^{-\tfrac{(p-1)}{2}\phi +2(5-p)B +2(6-p)C}\,\alpha'' + \frac{(\e^{-(p+1)A -\tfrac{p-3}{2}\phi + (7-p)B + (6-p)C} r^2)'}{r^2\e^{-(p+1)A + (p-3)B -(6-p)C +\phi}}\,\alpha' = m^2\e^{-2f}\,\alpha\,. 
}
\end{equation}
The brane charge, as measured at the horizon, is proportional to $\alpha'$, with a positive constant of proportionality:
\begin{equation}
Q = \alpha' r^2 e^{-(p+1)A-(p-3)\phi/2 + (7-p)B + (6-p)C} \,.
\end{equation}
Hence we obtain exactly the same no-go theorem for black brane solutions with anti-brane charge and regular $H$ field strength. The only solution with negative brane charge at the horizon must have a singular $H$ field strength given by
\be
\e^{-\phi}H^2 = \alpha^2 \e^{\phi -2(p+1)A + 2(p-6)(B+C) -2f} m^2\,.
\ee
where $\alpha$ evaluated at the horizon is non-zero.

The approach we followed necessarily involves brane solutions that are smeared over $6-p$ circle directions. This smearing leads to the simple ODE form of the equations of motion and to our analytic proof for the absence of black anti-branes. One might wonder whether a genuine localized anti-D3 brane solution could escape our `no-go' result. The answer is most likely negative, and this can be seen in two ways: First, if one smears nonextremal black holes along a flat direction, the horizon of the resulting solution is still smooth,\footnote{For example smearing black D3 branes gives a solution that is T-dual to the black D4 brane, etc.} and hence if a smooth localized anti-D3 black brane solution existed, a smooth smeared solution should exist as well, which contradicts our results and the results of \cite{Bena:2012ek}. Second, it was recently shown in \cite{Gautason:2013zw} that the analytic argument for the singularity of smeared antibranes can be extended (given some very sensible and weak assumptions) to localized antibranes. Since the same argument that proves that smeared antibranes are singular also proves the absence of smeared black anti-branes, we believe it will be quite straightforward to combine the arguments of \cite{Gautason:2013zw} to our arguments and prove that no localized black antibrane solution exists either.

\section{Discussion}\label{Spec}

Let us now recapitulate and further elaborate on the picture
presented in the Introduction.  The crucial question is what happens
if we put a D-brane with negative charge in a flux background with
positive D-brane charge. There is one very simple possibility,
initially mentioned in \cite{DeWolfe:2004qx} and further developed in
\cite{Blaback:2012nf}: the flux is attracted gravitationally and
electrically towards the anti-brane, it falls in, piles up, and
eventually a critical density of flux near the anti-brane is reached
such that annihilation takes place. Thus, the increased density of
flux should wipe out the would-be metastable state, and the decay
process then proceeds perturbatively without tunneling.

This scenario is supported by the fact that no black brane with
negative charge can exist in a background with positive charge in the
fluxes, as argued numerically in \cite{Bena:2012ek} and analytically
here. Indeed, if one adds a black brane with negative charge to the
flux background, the flux will be attracted towards the black brane,
fall in and finally be swallowed by the black brane. Eventually, as
flux of positive charge accumulates, the initial negative charge of
the black brane becomes zero. Thanks to gravity, however, the in-fall
of flux will continue. It only stops when the black brane has
collected a sufficient amount of positive charge to repel the flux,
and counteract further in-fall. Furthermore, as found in
\cite{Bena:2012ek}, and as one can no doubt also find numerically in
our system, the final charge of the black brane only depends on the
temperature of the black brane, so that it is completely blind to the value and
the sign of the charge of the original black brane we put in. Hence,
the presence of a black brane event horizon acts as a catalyst that
triggers brane-flux annihilation. Furthermore, this happens even if
the temperature of the black brane which we put in is parametrically
smaller than its  anti-branes charge, which indicates that the flux
clumping will still happen at zero temperature, when we add pure
anti-branes.

Interestingly, there exists an earlier example of the same phenomenon
in a different context. In \cite{Danielsson:2009gi} black holes in
backgrounds leading to Lifshitz scaling symmetries in the holographic
dual, were constructed and studied. It was shown that the mass and
charge of these black holes were not independent quantities. It was
argued, just as above, that the charge and mass automatically adjust
themselves in order to reach an equilibrium with the surrounding flux.
Furthermore, the solutions of \cite{Danielsson:2009gi} have nontrivial
$F_2$ and $H_3$ fluxes, and are very similar to ours; however, their
dilaton is locked to a constant value, so they are not, as they stand,
explicit solutions of string theory.

One can also try to obtain a microscopic counterpart to the macroscopic black brane picture that we have discussed above, using a model of black holes as collections of branes and anti-branes in thermal equilibrium with a gas of open strings (living on the branes). Such a picture  correctly reproduces the entropy of nonextremal black branes (see
for example \cite{Horowitz:1996ay,Danielsson:2001xe}). To form a black brane we start with a number of D branes and anti-D branes that come
together. Some of these D branes and anti-D branes might annihilate
against each other, creating a gas of open and closed strings. The final balance between the branes and the gas of strings is determined through thermodynamics. In the presence of flux, this balance will get modified, and our results indicate that the resulting mix will always have more branes than antibranes. This collection of branes, antibranes and open strings can also emit closed strings, and as this happens the temperature of the configuration goes down and more and more of the antibrane-brane pairs annihilate. This process corresponds to the Hawking radiation of the black brane, and can be quite slow if the black hole is large and the horizon has small curvature. However, if one considers adding just several antibranes to a region with flux, the corresponding black hole will have a Planck-sized horizon, and will evaporate in a time of order the Planck time. Hence, the brane-flux annihilation will be essentially instantaneous. 

In conclusion, we believe our results strengthen the intuition that antibranes in flux backgrounds do not give rise to metastable vacua, but rather annihilate immediately against the flux. Probably the most dramatic consequence of this is that the uplift of supersymmetric string vacua using anti D3-branes suffers from serious problems; it remains to be
seen if realistic cosmological models can nevertheless be constructed using perhaps other uplift mechanisms.

\section*{Acknowledgments} 
We would like to thank Alex Buchel, Oscar Dias and Stanislav Kuperstein for interesting discussions. The work of IB is supported in part by the ANR grant 08-JCJC-0001-0, and by the ERC Starting Independent Researcher Grant 240210 - String-QCD-BH. UD and  JB are supported by the Swedish Research Council (VR), and the G\"oran Gustafsson Foundation. TVR is supported by a Pegasus Marie Curie fellowship of the FWO. 

\bibliographystyle{utphysmodb}
\bibliography{anti}

\providecommand{\href}[2]{#2}\begingroup\raggedright\begin{thebibliography}{10}

\bibitem{Kachru:2003aw}
S.~Kachru, R.~Kallosh, A.~D. Linde and S.~P. Trivedi,  {\em {De Sitter vacua in
  string theory}}, Phys.Rev. {\bf D68} (2003) 046005
[\href{http://www.arXiv.org/abs/hep-th/0301240}{{\tt hep-th/0301240}}].

\bibitem{Kachru:2003sx}
S.~Kachru, R.~Kallosh, A.~D. Linde, J.~M. Maldacena, L.~P. McAllister {\em et
  al.},  {\em {Towards inflation in string theory}}, JCAP {\bf 0310} (2003) 013
[\href{http://www.arXiv.org/abs/hep-th/0308055}{{\tt hep-th/0308055}}].

\bibitem{Kachru:2002gs}
S.~Kachru, J.~Pearson and H.~L. Verlinde,  {\em {Brane / flux annihilation and
  the string dual of a nonsupersymmetric field theory}}, JHEP {\bf 0206} (2002)
  021
[\href{http://www.arXiv.org/abs/hep-th/0112197}{{\tt hep-th/0112197}}].

\bibitem{Bena:2011fc}
I.~Bena, A.~Puhm and B.~Vercnocke,  {\em {Metastable Supertubes and
  non-extremal Black Hole Microstates}}, JHEP {\bf 1204} (2012) 100
[\href{http://www.arXiv.org/abs/1109.5180}{{\tt 1109.5180}}].

\bibitem{Bena:2012zi}
I.~Bena, A.~Puhm and B.~Vercnocke,  {\em {Non-extremal Black Hole Microstates:
  Fuzzballs of Fire or Fuzzballs of Fuzz ?}}, JHEP {\bf 1212} (2012) 014
[\href{http://www.arXiv.org/abs/1208.3468}{{\tt 1208.3468}}].

\bibitem{Klebanov:2000hb}
I.~R. Klebanov and M.~J. Strassler,  {\em {Supergravity and a confining gauge
  theory: Duality cascades and chi SB resolution of naked singularities}}, JHEP
  {\bf 0008} (2000) 052 [\href{http://www.arXiv.org/abs/hep-th/0007191}{{\tt
  hep-th/0007191}}].

\bibitem{Bena:2009xk}
I.~Bena, M.~Grana and N.~Halmagyi,  {\em {On the Existence of Meta-stable Vacua
  in Klebanov-Strassler}}, JHEP {\bf 1009} (2010) 087
[\href{http://www.arXiv.org/abs/0912.3519}{{\tt 0912.3519}}].

\bibitem{Bena:2011wh}
I.~Bena, G.~Giecold, M.~Grana, N.~Halmagyi and S.~Massai,  {\em {The
  backreaction of anti-D3 branes on the Klebanov-Strassler geometry}},
\href{http://www.arXiv.org/abs/1106.6165}{{\tt 1106.6165}}.

\bibitem{Bena:2011hz}
I.~Bena, G.~Giecold, M.~Grana, N.~Halmagyi and S.~Massai,  {\em {On Metastable
  Vacua and the Warped Deformed Conifold: Analytic Results}},
\href{http://www.arXiv.org/abs/1102.2403}{{\tt 1102.2403}}.

\bibitem{Massai:2012jn}
S.~Massai,  {\em {A Comment on anti-brane singularities in warped throats}},
\href{http://www.arXiv.org/abs/1202.3789}{{\tt 1202.3789}}.

\bibitem{Bena:2012bk}
I.~Bena, M.~Grana, S.~Kuperstein and S.~Massai,  {\em {Anti-D3's - Singular to
  the Bitter End}},
\href{http://www.arXiv.org/abs/1206.6369}{{\tt 1206.6369}}.

\bibitem{Gautason:2013zw}
F.~F. Gautason, D.~Junghans and M.~Zagermann,  {\em {Cosmological Constant,
  Near Brane Behavior and Singularities}},
\href{http://www.arXiv.org/abs/1301.5647}{{\tt 1301.5647}}.

\bibitem{DeWolfe:2008zy}
O.~DeWolfe, S.~Kachru and M.~Mulligan,  {\em {A Gravity Dual of Metastable
  Dynamical Supersymmetry Breaking}}, Phys.Rev. {\bf D77} (2008) 065011
[\href{http://www.arXiv.org/abs/0801.1520}{{\tt 0801.1520}}].

\bibitem{McGuirk:2009xx}
P.~McGuirk, G.~Shiu and Y.~Sumitomo,  {\em {Non-supersymmetric infrared
  perturbations to the warped deformed conifold}}, Nucl.Phys. {\bf B842} (2011)
  383--413
[\href{http://www.arXiv.org/abs/0910.4581}{{\tt 0910.4581}}].

\bibitem{Bena:2012vz}
I.~Bena, M.~Grana, S.~Kuperstein and S.~Massai,  {\em {Polchinski-Strassler
  does not uplift Klebanov-Strassler}},
\href{http://www.arXiv.org/abs/1212.4828}{{\tt 1212.4828}}.

\bibitem{Polchinski:2000uf}
J.~Polchinski and M.~J. Strassler,  {\em {The String dual of a confining
  four-dimensional gauge theory}},
\href{http://www.arXiv.org/abs/hep-th/0003136}{{\tt hep-th/0003136}}.

\bibitem{Blaback:2010sj}
J.~Blaback, U.~H. Danielsson, D.~Junghans, T.~Van~Riet, T.~Wrase and
  M.~Zagermann,  {\em {Smeared versus localised sources in flux
  compactifications}}, JHEP {\bf 1012} (2010) 043
[\href{http://www.arXiv.org/abs/1009.1877}{{\tt 1009.1877}}].

\bibitem{Blaback:2011nz}
J.~Blaback, U.~H. Danielsson, D.~Junghans, T.~Van~Riet, T.~Wrase,  and
  M.~Zagermann,  {\em {The problematic backreaction of SUSY-breaking branes}},
  JHEP {\bf 1108} (2011) 105
[\href{http://www.arXiv.org/abs/1105.4879}{{\tt 1105.4879}}].

\bibitem{Blaback:2012nf}
J.~Blaback, U.~H. Danielsson and T.~Van~Riet,  {\em {Resolving anti-brane
  singularities through time-dependence}},
\href{http://www.arXiv.org/abs/1202.1132}{{\tt 1202.1132}}.

\bibitem{DeWolfe:2004qx}
O.~DeWolfe, S.~Kachru and H.~L. Verlinde,  {\em {The Giant inflaton}}, JHEP
  {\bf 0405} (2004) 017
[\href{http://www.arXiv.org/abs/hep-th/0403123}{{\tt hep-th/0403123}}].

\bibitem{Gubser:2000nd}
S.~S. Gubser,  {\em {Curvature singularities: The Good, the bad, and the
  naked}}, Adv.Theor.Math.Phys. {\bf 4} (2000) 679--745
[\href{http://www.arXiv.org/abs/hep-th/0002160}{{\tt hep-th/0002160}}].

\bibitem{Dymarsky:2011pm}
A.~Dymarsky,  {\em {On gravity dual of a metastable vacuum in
  Klebanov-Strassler theory}}, JHEP {\bf 1105} (2011) 053
[\href{http://www.arXiv.org/abs/1102.1734}{{\tt 1102.1734}}].

\bibitem{Horowitz:1995ta}
G.~T. Horowitz and R.~C. Myers,  {\em {The value of singularities}},
  Gen.Rel.Grav. {\bf 27} (1995) 915--919
[\href{http://www.arXiv.org/abs/gr-qc/9503062}{{\tt gr-qc/9503062}}].

\bibitem{Girardello:1999bd}
L.~Girardello, M.~Petrini, M.~Porrati and A.~Zaffaroni,  {\em {The Supergravity
  dual of N=1 superYang-Mills theory}}, Nucl.Phys. {\bf B569} (2000) 451--469
[\href{http://www.arXiv.org/abs/hep-th/9909047}{{\tt hep-th/9909047}}].

\bibitem{Freedman:2000xb}
D.~Z. Freedman and J.~A. Minahan,  {\em {Finite temperature effects in the
  supergravity dual of the N=1* gauge theory}}, JHEP {\bf 0101} (2001) 036
[\href{http://www.arXiv.org/abs/hep-th/0007250}{{\tt hep-th/0007250}}].

\bibitem{Klebanov:2000nc}
I.~R. Klebanov and A.~A. Tseytlin,  {\em {Gravity duals of supersymmetric SU(N)
  x SU(N+M) gauge theories}}, Nucl.Phys. {\bf B578} (2000) 123--138
[\href{http://www.arXiv.org/abs/hep-th/0002159}{{\tt hep-th/0002159}}].

\bibitem{Buchel:2000ch}
A.~Buchel,  {\em {Finite temperature resolution of the Klebanov-Tseytlin
  singularity}}, Nucl.Phys. {\bf B600} (2001) 219--234
[\href{http://www.arXiv.org/abs/hep-th/0011146}{{\tt hep-th/0011146}}].

\bibitem{Buchel:2001gw}
A.~Buchel, C.~Herzog, I.~R. Klebanov, L.~A. Pando~Zayas and A.~A. Tseytlin,
  {\em {Nonextremal gravity duals for fractional D-3 branes on the conifold}},
  JHEP {\bf 0104} (2001) 033
[\href{http://www.arXiv.org/abs/hep-th/0102105}{{\tt hep-th/0102105}}].

\bibitem{Gubser:2001ri}
S.~Gubser, C.~Herzog, I.~R. Klebanov and A.~A. Tseytlin,  {\em {Restoration of
  chiral symmetry: A Supergravity perspective}}, JHEP {\bf 0105} (2001) 028
[\href{http://www.arXiv.org/abs/hep-th/0102172}{{\tt hep-th/0102172}}].

\bibitem{Aharony:2007vg}
O.~Aharony, A.~Buchel and P.~Kerner,  {\em {The Black hole in the throat:
  Thermodynamics of strongly coupled cascading gauge theories}}, Phys.Rev. {\bf
  D76} (2007) 086005
[\href{http://www.arXiv.org/abs/0706.1768}{{\tt 0706.1768}}].

\bibitem{Buchel:2010wp}
A.~Buchel,  {\em {Chiral symmetry breaking in cascading gauge theory plasma}},
  Nucl.Phys. {\bf B847} (2011) 297--324
[\href{http://www.arXiv.org/abs/1012.2404}{{\tt 1012.2404}}].

\bibitem{Bena:2012ek}
I.~Bena, A.~Buchel and O.~J. Dias,  {\em {Horizons cannot save the Landscape}},
\href{http://www.arXiv.org/abs/1212.5162}{{\tt 1212.5162}}.

\bibitem{Blaback:2011pn}
J.~Blaback, U.~H. Danielsson, D.~Junghans, T.~Van~Riet, T.~Wrase,  and
  M.~Zagermann,  {\em {(Anti-)Brane backreaction beyond perturbation theory}},
  JHEP {\bf 1202} (2012) 025
[\href{http://www.arXiv.org/abs/1111.2605}{{\tt 1111.2605}}].

\bibitem{Bena:2012tx}
I.~Bena, D.~Junghans, S.~Kuperstein, T.~Van~Riet, T.~Wrase and M.~Zagermann,
  {\em {Persistent anti-brane singularities}}, JHEP {\bf 1210} (2012) 078
[\href{http://www.arXiv.org/abs/1205.1798}{{\tt 1205.1798}}].

\bibitem{Danielsson:2009gi}
U.~H. Danielsson and L.~Thorlacius,  {\em {Black holes in asymptotically
  Lifshitz spacetime}}, JHEP {\bf 0903} (2009) 070
[\href{http://www.arXiv.org/abs/0812.5088}{{\tt 0812.5088}}].

\bibitem{Janssen:1999sa}
B.~Janssen, P.~Meessen and T.~Ortin,  {\em {The D8-brane tied up: String and
  brane solutions in massive type IIA supergravity}}, Phys.Lett. {\bf B453}
  (1999) 229--236
[\href{http://www.arXiv.org/abs/hep-th/9901078}{{\tt hep-th/9901078}}].

\bibitem{Grana:2000jj}
M.~Grana and J.~Polchinski,  {\em {Supersymmetric three form flux perturbations
  on AdS(5)}}, Phys.Rev. {\bf D63} (2001) 026001
[\href{http://www.arXiv.org/abs/hep-th/0009211}{{\tt hep-th/0009211}}].

\bibitem{Blaback:2012mu}
J.~Blaback, B.~Janssen, T.~Van~Riet and B.~Vercnocke,  {\em {Fractional branes,
  warped compactifications and backreacted orientifold planes}}, JHEP {\bf
  1210} (2012) 139
[\href{http://www.arXiv.org/abs/1207.0814}{{\tt 1207.0814}}].

\bibitem{Horowitz:1996ay}
G.~T. Horowitz, J.~M. Maldacena and A.~Strominger,  {\em {Nonextremal black
  hole microstates and U duality}}, Phys.Lett. {\bf B383} (1996) 151--159
[\href{http://www.arXiv.org/abs/hep-th/9603109}{{\tt hep-th/9603109}}].

\bibitem{Danielsson:2001xe}
U.~H. Danielsson, A.~Guijosa and M.~Kruczenski,  {\em {Brane anti-brane systems
  at finite temperature and the entropy of black branes}}, JHEP {\bf 0109}
  (2001) 011
[\href{http://www.arXiv.org/abs/hep-th/0106201}{{\tt hep-th/0106201}}].

\end{thebibliography}\endgroup

\end{document}